\documentclass[review,10pt]{elsarticle}

\usepackage[version=4]{mhchem} 
\usepackage{textcomp}
\usepackage{graphicx}
\usepackage{subcaption} 
\usepackage{amsmath,amssymb}
\usepackage{bm}
\usepackage{array}
\usepackage{hyperref}
\usepackage{booktabs}
\usepackage{lineno}
\usepackage{xcolor}
\newcommand{\revised}[1]{#1}

\begin{document}

\begin{frontmatter}

\title{Melting Behavior and Phase Stability of CaO from Neural Network Potentials: a Molecular Dynamics Study}

\author[a]{Francesca Menescardi\corref{cor1}}
\ead{fmenesca@sissa.it} 

\author[a,b]{Stefano de Gironcoli}

\address[a]{Scuola Internazionale Superiore di Studi Avanzati (SISSA), via Bonomea 265, 34136 Trieste, Italy}
\address[b]{CNR-Istituto Officina dei Materiali (IOM), via Bonomea 265, 34136 Trieste, Italy}

\cortext[cor1]{Corresponding author}

\begin{abstract}
We investigate the melting behavior of calcium oxide (CaO) under extreme conditions, a problem that remains poorly constrained due to experimental limitations despite its relevance for geophysical and technological applications. We develop a Machine Learning Interatomic Potential (MLIP) for CaO with PANNA~2.0 and the LATTE descriptor, training it on a dataset of $\sim$12,000 configurations including solid, liquid, interfacial, and void-containing structures, extracted from ab-initio molecular dynamics data employing PBEsol exchange-correlation functional. We perform large-scale molecular dynamics simulations to compute the melting temperature at ambient pressure using both the void-nucleated melting (VNM) and two-phase coexistence (TPC) methods, obtaining $T_m=3055\pm11$~K and $T_m=2847\pm15$~K, respectively.\\
We calculate an enthalpy of fusion of $\Delta H_f\sim73$~kJ/mol, in agreement with thermodynamic assessments and ab initio calculations. We also reproduce the thermal expansion and obtain a volume increase of $\sim$29\% at Tm, consistent with the corresponding decrease in density extracted from spatially resolved number density profiles.
Finally, we calculate the high-pressure melting curve of CaO up to 20 GPa, providing one of the very few computational determinations of this quantity to date. The results confirm that the overheating ratio $\eta$ is not constant under pressure, increasing from 17\% at ambient pressure to 24\% at 20 GPa, confirming previous findings and ruling out the assumption of a fixed overheating ratio. Our results establish MLIP-based simulations as a robust and efficient framework for investigating phase stability in ionic oxides and provide new insight into the melting behavior of CaO under extreme conditions.
\end{abstract}


\end{frontmatter}

\section{Introduction}
Understanding high-temperature and high-pressure  (HT-HP) thermodynamic behavior of calcium oxide (CaO) is essential for the physics of planetary interiors, as it is a primary constituent of the silicate mantles of rocky planets\cite{Anderson1989}. Moreover, beyond its geophysical importance, CaO is a critical material in the development of refractory ceramics and industrial building materials, such as cement\cite{Taylor1997}. Despite its significance, establishing a precise melting point ($T_m$) for CaO is still a major challenge, since the experimental values range widely from 2843~K to 3223~K\cite{Liang2018}. This discrepancy is largely attributed to the material's high vapor pressure and very high reactivity, in particular its tendency to form calcium hydroxide (Ca(OH)$_2$) or calcium carbonate (CaCO$_3$) when put in contact with H$_2$O or CO$_2$. Moreover, while recent laser-heating studies\cite{Manara2014} have helped to certify that the melting temperature for this material is likely well above 3000~K, the high pressure melting curve has remained largely unexplored experimentally. \\
In our previous research\cite{menescardi2024} we employed atomistic simulations to investigate these properties, using both Born-Mayer-Huggins (BMH) empirical potential and \textit{ab-initio} molecular dynamics (AIMD). A significant finding of that work, besides the calculation of the melting temperature employing different techniques, was that the overheating ratio ($\eta$) is not constant under pressure, which contradicts previous assumptions\cite{Sun2010} that $T_m$ could be reliably estimated by scaling the thermal instability limit by a fixed factor. However, the empirical BMH potential employed in that work can lack the necessary quantum accuracy and transferability required to map a full phase diagram under these extreme conditions. Conversely, while AIMD provides the required precision, its computational cost is several orders of magnitude higher than classical MD, making it prohibitive for the large systems and long simulation times needed for accurate melting curve calculations.\\
\revised{Machine Learning Interatomic Potentials (MLIPs) have emerged as a solution to bridge this gap, becoming an established tool in computational materials science~\cite{behler2016,unke2021,butler2018}, achieving near-ab-initio accuracy at a computational cost comparable to that of classical empirical potentials. This approach has enabled large-scale molecular dynamics simulations - up to tens of thousands of atoms over nanosecond timescales - and has been shown to achieve chemical accuracy across a wide range of systems~\cite{deringer2017,artrith2012,jain2021,wang2024construction}.}\\
For this reason, the present study develops and validates a novel MLIP for CaO, utilizing the PANNA 2.0 framework~\cite{pellegrini2023panna2} and the advanced LATTE descriptor~\cite{pellegrini2024latte}. We employ a multilayer perceptron (MLP) architecture, chosen for its computational efficiency in large-scale MD production runs. The primary objectives of this work are to improve the results previously obtained for the melting temperature of this system and to rigorously calculate the high-pressure melting curve up to 20~GPa, resolving the thermodynamic inconsistencies observed in previous empirical models and validating the results. By training the potential on a diverse dataset of solid, liquid, defected and high-pressure structures, we aim to provide a more robust certification of the CaO phase diagram.
\section{Computational Methods}

\subsection{Neural Network Potential and Dataset}
In recent years, machine learned interatomic potentials (MLIPs) have been used effectively to try to achieve \textit{ab-initio} accuracy at a cost comparable with that of empirical potentials. In fact, while \textit{ab-initio} Molecular Dynamics (AIMD) is highly accurate, its computational cost remains several orders of magnitude higher than Classical Molecular Dynamics (CMD). Conversely, empirical force fields are computationally fast but can lack of accuracy and transferability, particularly under the extreme pressure and temperature conditions required to map a full HP-HT phase diagram. To bridge this crucial gap, we employ a Machine Learning Interatomic Potential (MLIP) approach, aiming to achieve \textit{ab-initio} accuracy at a computational cost comparable with that of an empirical potential.\\
For this purpose, we employ the PANNA\cite{lot2020panna,pellegrini2023panna2} code, a comprehensive software package for generating MLIPs based on local atomic descriptors and multilayer perceptrons (MLPs). The selection of the MLP architecture is motivated by computational efficiency, as MLPs are faster for inference than more recent Graph Neural Network (GNN) architectures, making them suitable for the extensive and large-scale MD production runs required for melting studies. PANNA adopts a modified version of the method originally proposed by Behler and Parrinello \cite{behler2007}. As for the majority of empirical potentials, in this framework the total energy of a system is expressed as the sum of the single atoms energies:
\begin{equation}
    E = \sum_{s}\sum_{i\in s} E_{s}(\{d_i\}) 
\end{equation}
where $s$ is the atomic species and \textit{i} is an index that labels atoms in the unit cell. Each atomic energy term is expressed as a function of the local environment, which is encoded in a tensor called the \textit{descriptor} ($\{d_i\}$). The descriptor, invariant under translations, rotations and permutations of equivalent atoms, is processed through the machine learning architecture to obtain the predicted element-independent atomic energies $E_s(\{d\})$. In this work, we employ LATTE descriptor\cite{pellegrini2024latte}, which is based on Cartesian tensor contractions and is explicitly designed for the efficient construction of a variable number of many-body terms with learnable parameters. Each tensor component is constructed as
\begin{equation}
A^{\alpha_1 \dots \alpha_p}_i = \sum_{j \in \mathcal{N}(i)} \sigma_\nu(s_i, s_j)\, f_{\nu,s_i}(|\vec{r}_{ij}|)\, \hat{r}^{\alpha_1}_{ij} \dots \hat{r}^{\alpha_p}_{ij},
\end{equation}
where $\sigma_\nu$ is a learnable species-dependent weight, $f_{\nu,s_i}$ is a radial function, and $\hat{r}_{ij}$ is the unit vector pointing from atom $i$ to its neighbor $j$. The multi-index $\alpha_1 \dots \alpha_p$ encodes angular information up to a given body order $p$, and the sum runs over the neighbors $\mathcal{N}(i)$.

The radial function $f_{\nu,s_i}(r)$ is defined as:
\begin{equation}
f_{\nu,s_i}(r) = \frac{1}{r^3_{\nu,s_i}} \left[ \text{ReLU} \left( 1 - \left( \frac{r - r_{\nu,s_i}}{w_{\nu,s_i}} \right)^2 \right) \right]^3,
\end{equation}
where $r_{\nu,s_i}$ and $w_{\nu,s_i}$ are learnable parameters, and $\text{ReLU}(x) = \max(0, x)$ is the rectified linear unit. This function ensures smooth cutoff behavior and emphasizes neighbors at specific distances. This use of a localized radial function helps the subsequent MLP model to fit single features more accurately, leading to better generalization.

From $N$ such tensorial features, the final descriptor is constructed as an $N{+}1$-body invariant by taking a product of tensors and contracting over angular indices:
\begin{equation}
B_{i,u} = \sum_{\alpha_1, \dots, \alpha_m} \prod_{k=1}^{N} A^{\alpha_1^k \dots \alpha_{p_k}^k}_{i,u_k},
\end{equation}
where it is implied that each of the indices $\alpha_1, \dots, \alpha_m$ appears exactly twice among the multi-indices $\alpha_j^k$, meaning that every tensor component is contracted with exactly one other component. The indices $u_k$ label the learnable parameters for each tensor component $A_{i,u_k}$, while $u$ indexes the overall descriptor bin $B_{i,u}$.

The 2-body term is a special case of this formulation, in which $B_{i,u}$ corresponds directly to a scalar $A_{i,u}$ with no angular indices. In this case, the descriptor simply reduces to a radial function of neighbor distances and species identities.

As for the shape of the NN, PANNA adopts MLP-type networks, for which the general equation is:
\begin{equation}
    a_i^l = \sigma\left( \sum_{j=1}^{n_{l-1}} w_{ij}^l a_j^{l-1} + b_i^l \right)
\end{equation}
where $a_i$ is a node of layer, $l,w$ and $b$ are the parameters of the network (i.e. weights and biases), $\sigma$ is the activation function and $n_l$ is the number of nodes in the layer $l$. The activation function $\sigma$ is a nonlinear function and can be chosen to be Gaussian, rectified linear unit (ReLU) or hyperbolic tangent; only the last layer has a linear activation function, as it is typically used for the output. \\
Given an architecture, the model is then trained (i.e. optimized) minimizing the loss function, which represents the deviation of the values in energies and forces predicted from the ones computed from \textit{first-principles}. \\

\subsubsection{First principles calculations}
 To build the dataset, we carried out AIMD simulations using GPU-enabled Quantum ESPRESSOv6.8 package\cite{Giannozzi2009, Giannozzi2017, Giannozzi2020}. We employ standard norm-conserving pseudopotentials\cite{Troullier1991}, with a wave function cutoff of 70 Ry. We employ the PBEsol\cite{Perdew2008} exchange-correlation functional, which is known\cite{Csonka2009} to predict fairly well the equations of state of many inorganic materials and to perform in agreement with experiments on several structural and thermodynamics properties such as density and enthalpies of formation.\\
 \subsubsection{Dataset composition}
 To craft an appropriate dataset to train the neural network, we sampled all the configurations that were likely to be explored during the entire range of MD simulations. The bulk of the dataset consists of solid and liquid structures extracted from AIMD simulations of 3$\times$3$\times$3 CaO supercells at different temperatures and ambient pressure. The supercells are replicas of the conventional 8 atoms $fcc$ cubic cell of CaO. Since the melting temperature of the system was predicted to be $\sim$3000 K\cite{menescardi2024}, we chose the temperature range for the solid structure to be 300 K - 3200 K, covering the entire range every 500 K. Similarly, the liquid portion of the dataset covers temperatures from 2000 K to 6000 K, with a step of 500 K. Combined, solid and liquid structures at different temperatures make up for $\sim$70\% of the entire dataset. \\
 The remaining $\sim$30\% covers the remaining simulation scenarios, which are explained in more details in section \ref{sec:results}. To model the solid-liquid interface to accurately perform the two-phase solid-liquid technique, we filled $\sim$15\% of the dataset with 2$\times$2$\times$4 supercell of half-solid half-liquid structures of CaO, equilibrated at different temperatures around the supposed melting temperature (i.e. 2900 K, 3000 K and 3100 K). To accurately model the void-melting technique, instead, $\sim$8\% of the dataset is composed by 3$\times$3$\times$3 solid "voided" structures, with a hole at the center of the cell increasing in dimension from 1\% to 40\% in volume. Lastly, the remaining $\sim$7\% of the dataset is composed by expanded or compressed solid structures to account for pressure or volume changes up to 20 GPa. The dataset breakdown is summarized in table \ref{tab:dataset}.    

\begin{table*}\begin{center}
\renewcommand{\arraystretch}{1.3}
\begin{tabular}{cccccc}
\hline\hline
\textbf{Structure Type} & \textbf{Range} & \textbf{Proportion} \\
\hline
Solid   & 300 K - 3200 K & $\sim$35\%\\
Liquid   & 2000 K - 6000 K & $\sim$35\%\\
Interface   & 2900 K, 3000 K, 3100 K & $\sim$15\%\\
Voided   & 1\% - 40\% in volume & $\sim$8\%\\
EOS   & $-$20 GPa - 20 GPa & $\sim$7\%\\
\hline\hline
\end{tabular}
\caption{Composition of the dataset divided by types of structures.}\label{tab:dataset}
\end{center}\end{table*}

The dataset is composed of $\sim$12,000 structural configurations, which were split into a training set containing $\sim$75\% of the structures, a validation set containing $\sim$15\% of the structures and a test set containing $\sim$10\% of the structures. 

\subsubsection{Training details}

\begin{figure}[h]
\centering
\begin{subfigure}{0.49\textwidth}
    \includegraphics[width=\linewidth]{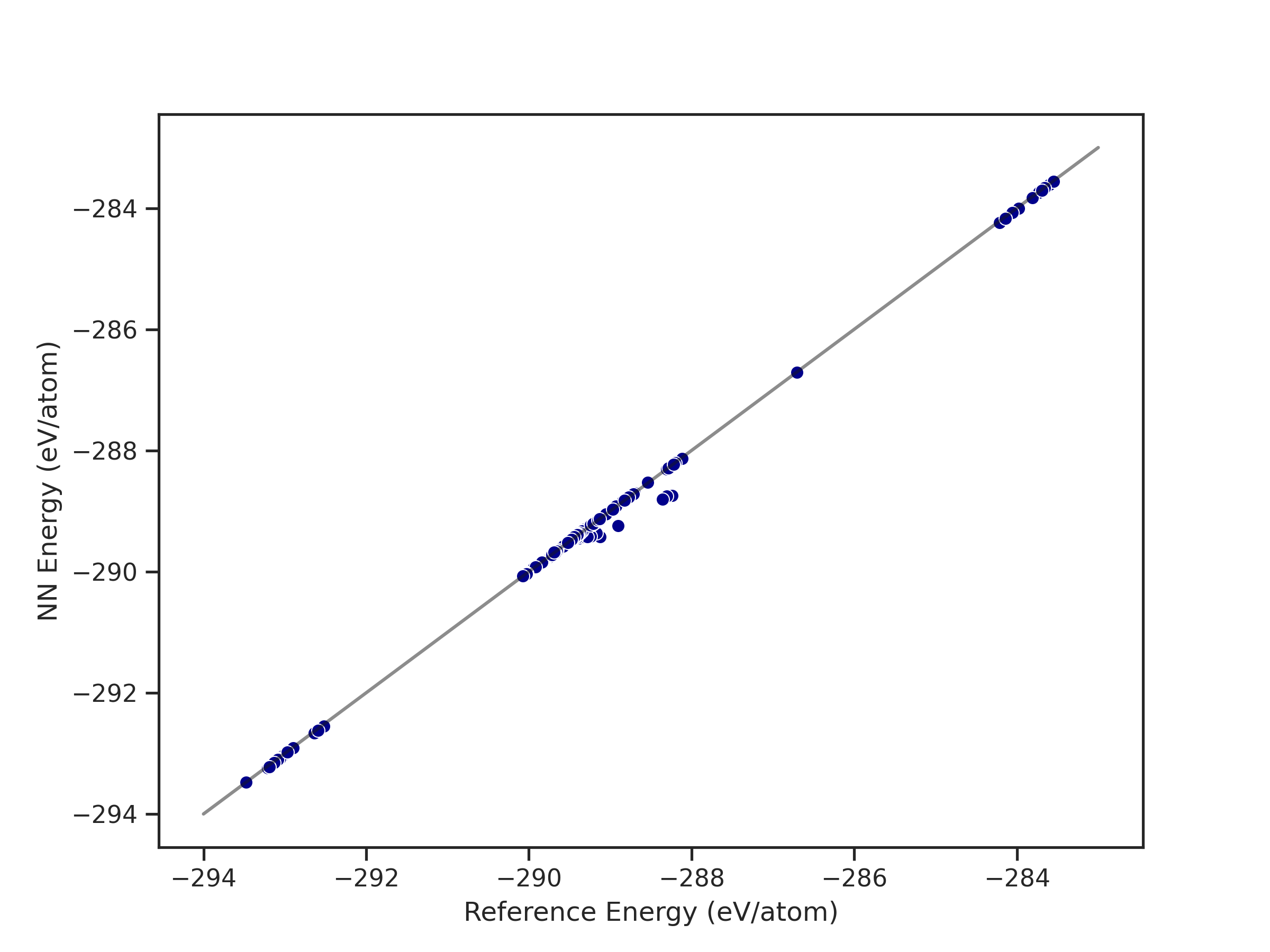}
    \caption{}
    \label{fig:parity_en}
\end{subfigure}
\hfill
\begin{subfigure}{0.49\textwidth}
    \includegraphics[width=\linewidth]{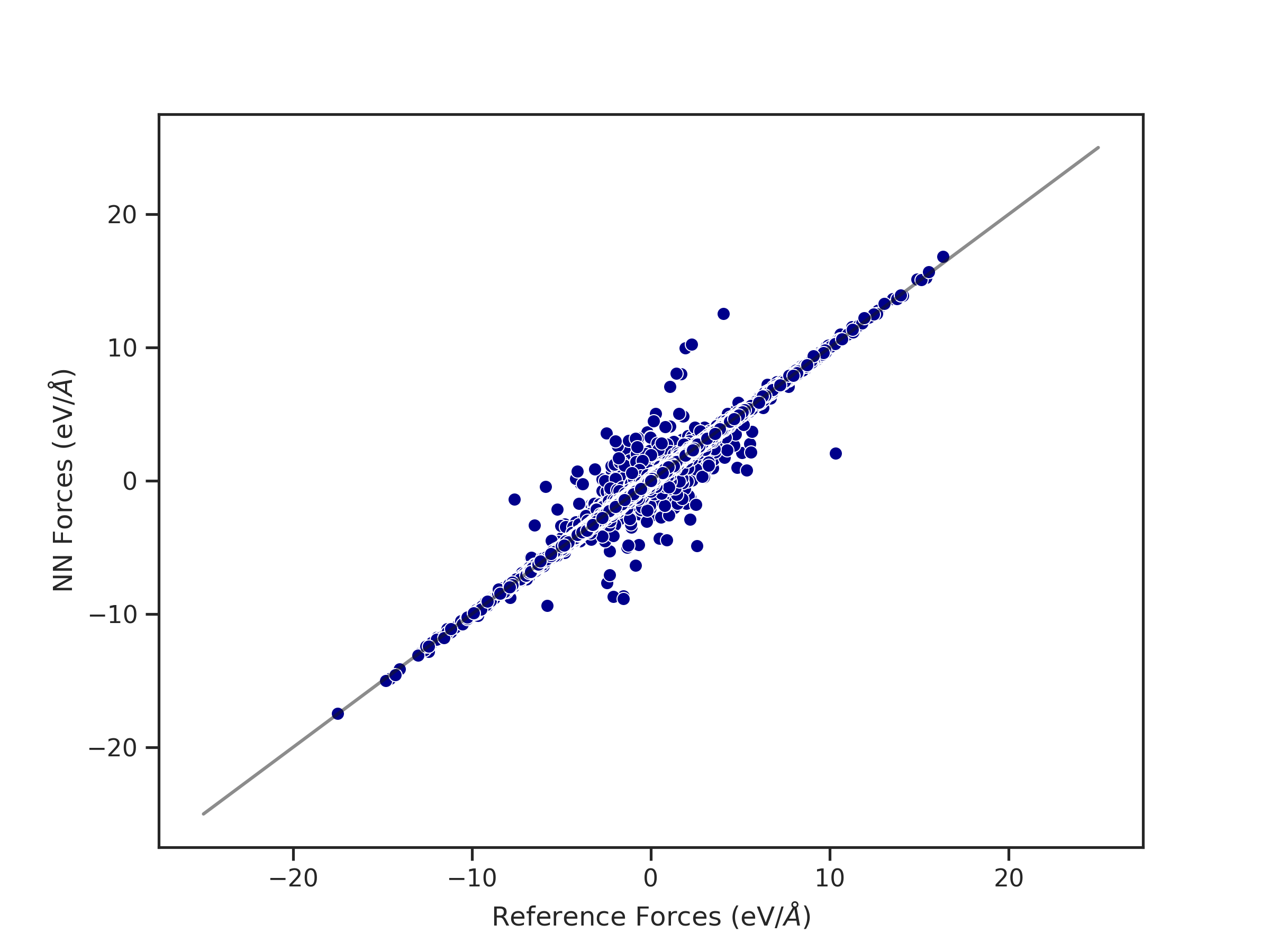}
    \caption{}
    \label{fig:parity_for}
\end{subfigure}
\caption{Parity plot validating the energy (a) and forces (b)}
\label{fig:parity}
\end{figure}

The potential uses a fully-connected, deep, feed-forward Neural Network (FFNN) architecture consisting of three layers, 256:124:1. The two hidden layers employ non linear activation functions (Gaussian) while the final output layer uses a linear activation function to output the predicted atomic energy and forces contributions. The influence of neighboring atoms is restricted by a fixed radial cutoff of $R_c$ = 5.0 \AA.\\
The fidelity of the potential relies heavily on the quality of the local atomic descriptor. The full descriptor shape used is defined by the string notation:
\begin{equation}
    d_i=100,-:200,i,i:200,ij,ij:200,ijk,ijk:200,i,j,ij
\end{equation}
The entire descriptor comprises five different tensor contraction terms separated by semicolons, summing to 900 features. The indices represent the number of features in the descriptor for each body-order complexity: a 2-body term ($-$), three 3-body terms ($i,i$ and $ij,ij$ and $ijk,ijk$) and a 4-body term ($i,j,ij$). By defining complex tensor contractions, the descriptor gains high expressiveness, essential for capturing complex short-range forces and structural changes inherent to high-temperature and high-pressure simulations, without loosing the invariance ensured by the contraction method. \\
The learning rate was scheduled to range from $10^{-4}$ to $10^{-7}$, with an exponential decay and a final quench to favor convergence, and involved 3000 epochs. The training performance was verified in terms of the Mean Absolute Error (MAE) on energy and forces, with the final values being MAE$_{energy}<5$ meV/atom and MAE$_{forces}<45$ meV/\AA. The corresponding parity plots for energy and forces are shown in fig\ref{fig:parity_en} and fig\ref{fig:parity_for} respectively.

The apparent scatter in the forces parity plot is mainly associated with a limited subset of high-energy configurations, such as high-temperature liquid states and structures containing voids. These configurations produce large force magnitudes that disproportionately affect the visual appearance of the plot, even though they account for less than 5\% of the dataset. The remaining configurations, which constitute the majority, show much closer agreement with the reference calculations and are tightly distributed around the parity line. \revised{To further contextualize the reported MAE beyond this visual inspection, we computed the standard deviation of the force components across the full training and test sets, obtaining $\sigma_{train} = 1.555$~eV/\text{\AA} and $\sigma_{test} = 1.559$~eV/\text{\AA}. The force MAE is therefore only $\sim$2.9\% of this intrinsic spread in both sets, indicating that the force errors are small relative to the typical scale of variation of the forces sampled in the dataset.}

\subsubsection{Molecular Dynamics}
To perform all the MD simulations we employ the LAMMPS code~\cite{Plimpton1995,Thompson2022}. Periodic boundary conditions are applied in all three directions, supercell dimensions and total simulation time vary depending on the different techniques that we use (i.e. the void-nucleated melting or the two-phase coexistence method, see Section\ref{sec:results}), and all simulations are performed in the isobaric (NPT) ensemble, employing the Nosé-Hoover barostat and thermostat~\cite{Hoover1985}, except otherwise stated. We always use a time step of 0.5~fs for a minimum of 600,000 steps which makes the minimum total simulation time of 300 ps. With this MLIP, we are able to run simulations of very large supercells, which, depending on the technique employed, contain a minimum of 10,648 atoms and a maximum of 17,280 atoms.
\section{Results}\label{sec:results}

\subsection{Melting temperature of CaO}
The melting point ($T_m$) is one of the most fundamental yet challenging properties to determine computationally, especially for high-temperature refractory materials like CaO\cite{Hong2022,Alvares2020}. From a thermodynamic perspective, the $T_m$ of a crystal is rigorously defined as the temperature at which the molar Gibbs free energies of the solid phase and the liquid phase are equal at any given pressure. However, accurately locating this thermodynamic equilibrium using MD simulations presents the significant hurdle of superheating\cite{Zhang2012}, which is defined as heating a perfect crystal beyond $T_m$ without the onset of melting. MD simulations performed on ideal, defectless crystals inevitably lead to homogeneous melting at a significant higher limit known as the mechanical melting point, or thermal instability temperature\cite{Belonoshko2006}, which is the temperature above which the structural integrity of the crystal lattice collapses. This limit of superheating ($T_s$) is triggered when the internal energy of the solid is sufficient to cause the spontaneous nucleation of liquid regions through accumulating local vibrational and elastic lattice instabilities\cite{Jin2001,Belonoshko2006} and it is significantly higher than the actual melting temperature of the system. In real systems, in fact, lattice defects, free surfaces and grain boundaries, which are statistically very common, assist in the nucleation of the melting at temperatures very close to $T_m$, which are usually 20-25\% lower than $T_s$\cite{Jin2001}.  \\
To overcome this intrinsic superheating artifact, computational strategies rely on heterogeneous nucleation techniques, of which we selected two strategies: the \textit{void-nucleated melting} and the \textit{two-phase coexistence} technique. \revised{A third approach relies instead on the direct calculation of the solid-liquid free-energy difference via thermodynamic integration\cite{Kirkwood1935,FrenkelSmit2023} or non-equilibrium free-energy calculations\cite{Freitas2016,Leite2019}, widespread techniques that have proven affordable in combination with MLIP-based molecular dynamics and that will be the subject of future works on this subject.}

\subsubsection{Void-Nucleated Melting Technique}
\label{sec:VNM}

The Void-Nucleated Melting (VNM) technique is a crucial direct simulation method used in molecular dynamics to determine the thermodynamic melting temperature ($T_m$) of a crystal\cite{Lutsko1989}. VNM overcomes the artifact of superheating by intentionally introducing a crystal defect, such as a spherical or cubic void, into the perfect lattice\cite{wang2024,zou2020}. These extrinsic defects provide favorable conditions for the formation and growth of the liquid phase, fundamentally initiating heterogeneous nucleation by lowering the local free energy barrier required for the liquid phase to form\cite{Zhang2012}. For simple systems like the Lennard-Jones fluid or Argon, VNM is categorized as relatively easy to apply and generally accurate, reproducing experimental values reasonably well\cite{Zhang2012}. Its utility has also been shown for refractory oxides like CaO\cite{menescardi2024}. 
The melting procedure typically involves heating the system and monitoring sharp discontinuities in thermodynamic properties like enthalpy or volume to locate the apparent melting point ($T_c$). A void is created in the relaxed crystal structure of an 11$\times$11$\times$11 supercell of the conventional $fcc$ cell (10648 atoms in total) by removing $n$ atoms from the center to create the nucleation point. The system is then gradually heated up in a $NPT$ ensemble at ambient pressure and $\sim 10^{12}$ K/s heating rate up until it melts completely. The simulation is then repeated consistently increasing the dimension of the void from $n=0$, which represents the defect-less crystal and outputs the thermal instability temperature $T_s$, up to when the void dimension gets so big that causes the premature collapse of the crystal structure, which in our case corresponds to $\sim$45\% in volume. Given that the apparent melting point depends on void density, the apparent melting temperature lowers gradually as the void increases with respect to the supercell volume, until $T_c$ stabilizes on a temperature plateau.The average $T_c$ at the plateau represents the true thermodynamic melting point, because it is independent of the void size.

\begin{figure}[h]
\centering
    \includegraphics[width=\linewidth]{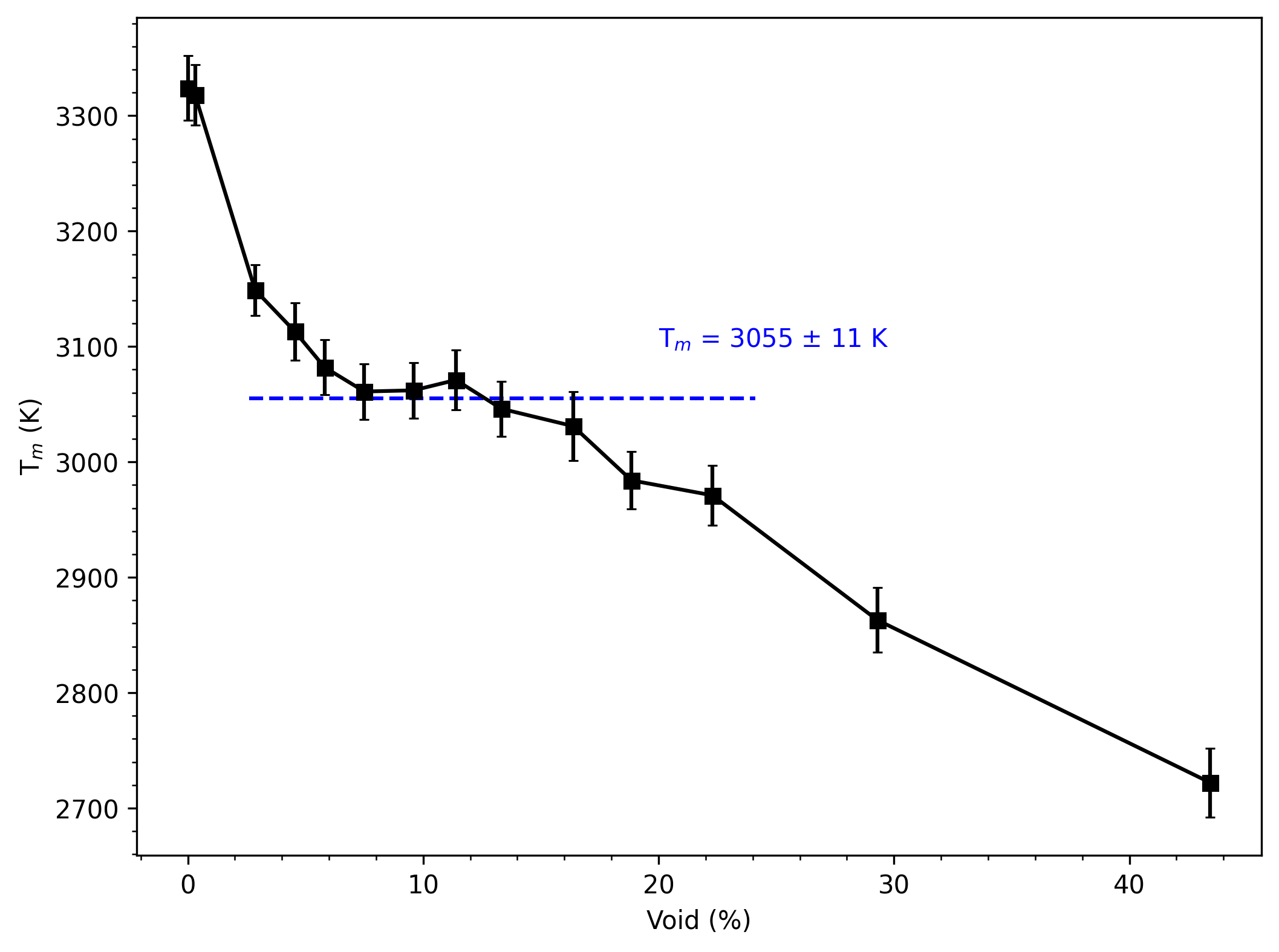}
    \caption{Apparent melting temperature of crystal CaO as a function of the defect dimensions (in \% of the supercell volume). \revised{For each void size, the melting transition was identified from the discontinuity in cell volume, and error bars represent the standard deviation of the temperature within the identified transition window.} The plateau is reached at $T_m = 3055 \pm 11$~K, which represents the melting temperature of the solid; \revised{the reported uncertainty on this value is obtained by propagating in quadrature the individual uncertainties of the plateau points ($\sigma_{T_m} = \sqrt{\sum_i \sigma_i^2}/N$), divided by the number of points averaged.}}
\label{fig:void}
\end{figure}

Figure\ref{fig:void} illustrates the apparent melting temperature $T_c$ as a function of the dimensions of the defect as a \% in volume of the 11$\times$11$\times$11 supercell. $T_c$ slowly decreases as the void dimension increases, up until $\sim$8\% in volume, where the decreasing trend flattens in a plateau up until $\sim$17\% in volume. Further increasing the void dimensions causes more and more instability in the crystal structure, resulting in the collapse of the supercell on its own and an unreliably low apparent melting temperature. Averaging the $T_c$ values obtained on the plateau line, we obtain $T_m=3055\pm11$ K.
Although the VNM technique faces significant limitations, principally because of its reliance on empirical observation without a specified thermodynamic basis\cite{Zhang2012,zou2020}, the melting temperature falls fairly close to the results of $T_m=3222\pm25$ K and $T_m=3160\pm10$ K obtained in the most recent experiments\cite{Manara2014,Bgasheva2021}. Compared to the previous computational studies, the value obtained is perfectly in line with results obtained both employing an empirical potential ($T_m=3066\pm12$~K\cite{menescardi2024}) and a MLIP ($T_m=3057$~K\cite{Lee2022}) confirming the reliability and the robustness of the MLIP we trained and certifying a melting temperature of CaO above 3000~K.

\subsubsection{Two-Phase Coexistence Method}
\label{sec:TPC}
The second method we applied to calculate the melting temperature of CaO is the Two-Phase Coexistence (TPC) method. This technique was first employed to obtain the melting curve of aluminium\cite{Morris1994} and since then it has been widely used as a computational strategy to obtain the $T_m$ of a large variety of materials\cite{Agrawal2003,Hong2013,Wang2023}. The principle behind the technique is intuitive: one half of a big supercell is maintained solid, while the other half is heated to very high temperature to ensure the complete melting, and then cooled down to a temperature close to the supposed $T_m$ of the material; once created this system in which solid and liquid phases coexist, MD simulations at different temperatures are performed up until the solid and the liquid phases are at equilibrium, temperature which corresponds to the melting temperature of the crystal. \\
In this work, we applied the procedure proposed by Asadi et al.\cite{asadi2015}, which consists in constructing a $m\times n\times l$ supercell where the $l$ dimension, normal to the solid-liquid interface, is much longer than the other two directions. 
\revised{First, the perfect crystal supercell is equilibrated under an isothermal-isobaric ($NPT$) ensemble at ambient pressure and a temperature of 3000~K, which corresponds to an estimated melting temperature for this system, for 50~ps. This step allows the in-plane lattice parameter to reach its correct thermally-expanded equilibrium value at the target temperature. We verified that this in-plane parameter converges to a stable plateau within approximately 1-1.5~ps, well within the total duration of this equilibration stage, with the pressure fluctuating around zero, indicating mechanical equilibrium.} The central half of the simulation box is then melted at very high temperature in a canonical ($NVT$) ensemble employing a Nosé-Hoover thermostat\cite{Hoover1985}, while the other half is fixed. The liquid half of the simulation box is then equilibrated for 50 ps in a isothermal-isobaric ($NPT$) ensemble at 3000~K and ambient pressure. In this simulation the $z$ direction of the simulation box, corresponding to the direction normal to the solid-liquid interface, is allowed to relax to account for the changes in volume, while the solid portion of the supercell is kept fixed. To minimize the pressure in all directions, the solid half is then unfixed, and the entire simulation box is allowed to relax in the normal direction in a $NPT$ ensemble for 0.5 ns at the estimated melting temperature. Finally, the refined $T_m$ value is calculated by running a longer simulation (2.5 ns) in an iso-enthalpic ($NPH$) ensemble, always allowing the system to relax in the normal direction. The refined temperature is calculated as the average temperature of the system during this last MD simulation and, if it is significantly different from the estimated melting point, the process is repeated using the new refined melting temperature as the estimated melting point up until convergence is achieved.\\ 
To test if the dimensions of the system play a role in the determination of the melting temperature, we tested five different combinations of $m\times n\times l$ as a supercell. The results are shown in table \ref{tab:two-phase}.

\begin{table*}\begin{center}
\renewcommand{\arraystretch}{1.3}
\begin{tabular}{cccccc}
\hline\hline
\textbf{$m\times n\times l$} & \textbf{\# atoms} & \textbf{$T_m$ (K)} \\
\hline
$6\times 6\times 12$   & 3,456 & N.A. \\
$6\times 6\times 20$   & 5,760 & 2,864 $\pm$ 32 \\
$6\times 6\times 40$   & 11,520 & 2,837 $\pm$ 19\\
$6\times 6\times 60$   & 17,280 & 2,847 $\pm$ 15\\
$8\times 8\times 30$   & 15,360 & 2,825 $\pm$ 17\\
\hline\hline
\end{tabular}
\caption{Melting temperature calculated with TPC technique for different supercell dimensions}\label{tab:two-phase}
\end{center}\end{table*}

\begin{figure}[h]
\centering
    \includegraphics[width=\linewidth]{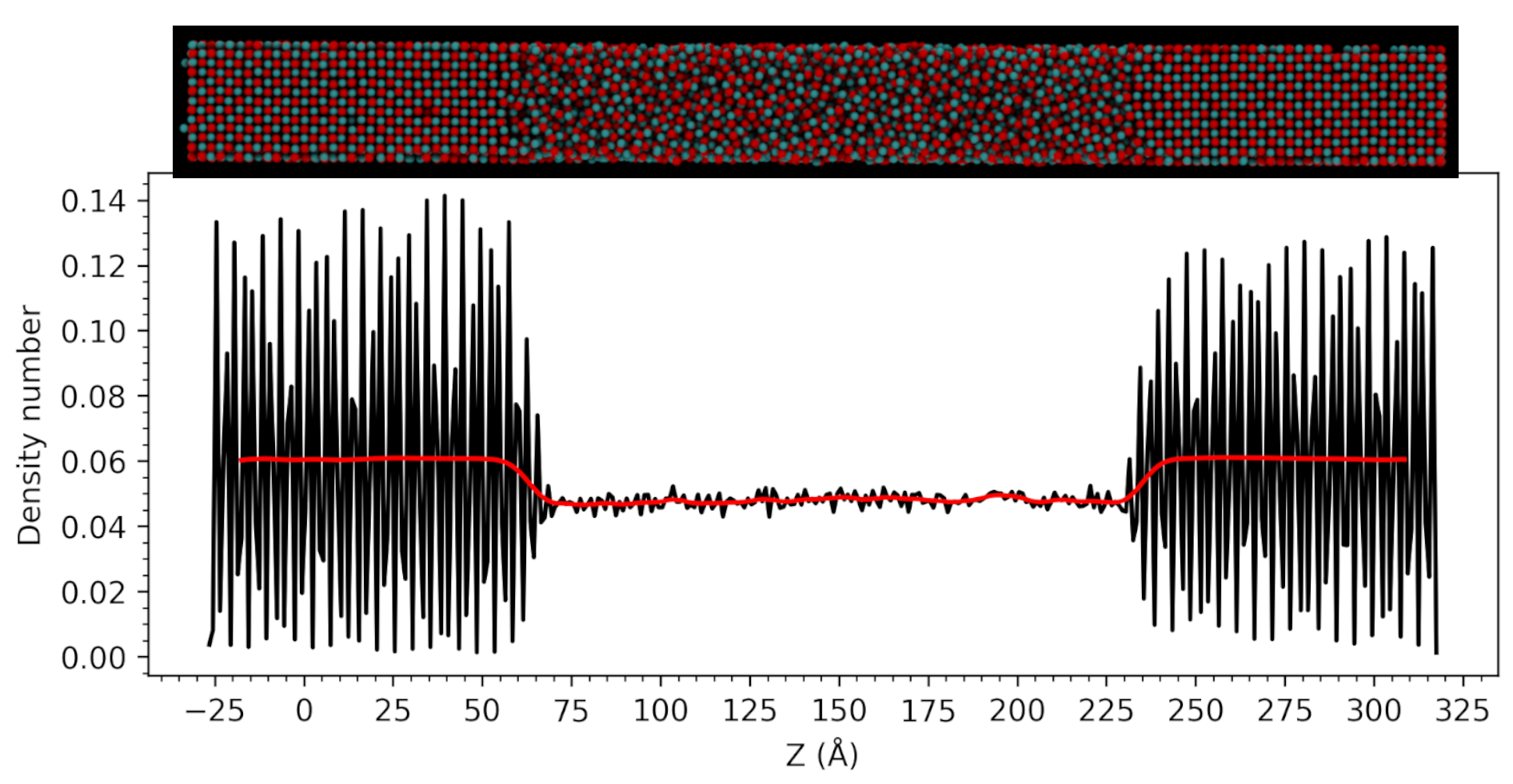}
    \caption{Snapshot of the 6$\times$6$\times$60 supercell, visualized with VMD\cite{Humphrey1996}, where the red spheres represent the oxygen atoms while the blue spheres represent the calcium atoms. Below, the corresponding number density, calculated as an average over 10 ps with VMD Density Profile Tool\cite{giorgino2014}. The red line represents the spatially averaged density profile, which highlights that the crystal phase has a higher density than the liquid phase.}
\label{fig:dens_num}
\end{figure}

With the exception of the $6\times 6\times 12$ supercell, which contained too few atoms to obtain a reliable melting temperature, there is no significant variation in the $T_m$ results, so we conclude that the calculated $T_m$ is size-independent. Even keeping the most accurate result of 2,847 $\pm$ 15~K obtained with the 6$\times$6$\times$60 supercell, the $T_m$ calculated with the TPC method in this work is lower than the latest experimental results~\cite{Manara2014,Bgasheva2021}. \revised{While anisotropic interfacial pressure arising from the solid-liquid interface stress tensor is a recognized source of finite-size artifacts in two-phase coexistence simulations~\cite{zou2020}, this discrepancy is more plausibly related to specific factors affecting the accuracy of the underlying interatomic potential~\cite{wang2024}. For this reason, neither VNM nor TPC can be regarded as intrinsically more reliable than the other on the sole basis of agreement with experiment, and both $T_m$ estimates are  used as reference throughout this work.}\\
Figure \ref{fig:dens_num} shows a snapshot of the 6$\times$6$\times$60 simulation box at the ambient-pressure melting point, compared to the density number profile averaged over the last 10 ps of the NPH simulation. It can be observed that the solid-liquid interface extends over 4-5 atomic planes (about 10 \AA) between the liquid and the solid portions of the supercell and the density profile of the liquid does not show any sign of periodic arrangement. The figure also shows that, at the melting temperature, $\rho_{crystal}>\rho_{liquid}$, as highlighted by the spatially averaged density profile, represented by a red solid line. Specifically, the density of the solid phase at melting temperature is on average $\sim$27\% higher than the density of the liquid phase.

\subsection{Enthalpy of Fusion}
\begin{figure}[h!]
\centering
    \includegraphics[width=\linewidth]{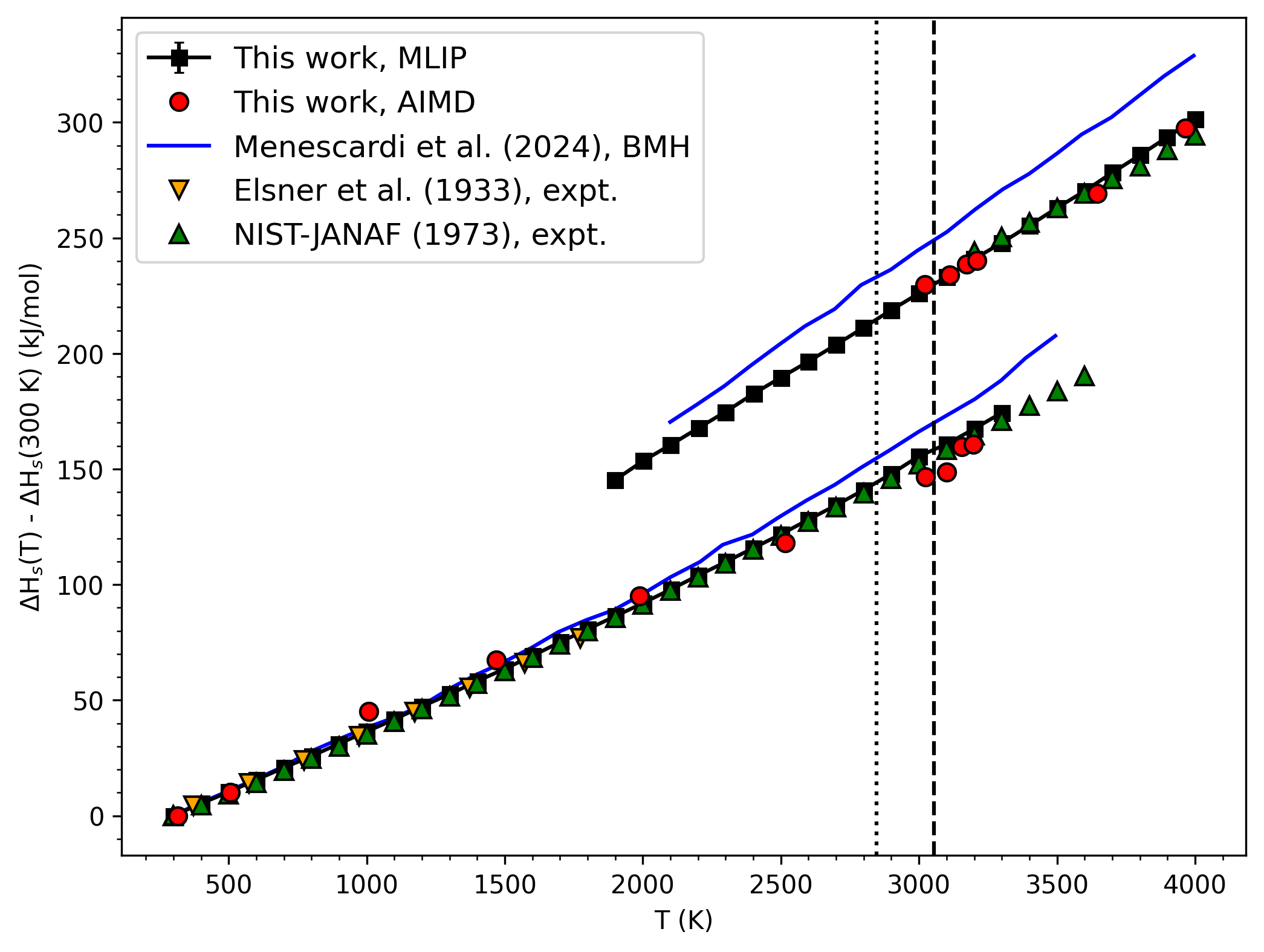}
    \caption{Enthalpy difference relative to the enthalpy of the solid phase at 300~K ($H_{S,300}^{0}$) as a function of the temperature. \revised{Error bars represent the standard deviation of the enthalpy during the production phase of each simulation (last 40,000 of 50,000 steps).} The lower branch is referred to the crystal phase, while the upper branch is related to the liquid phase. The vertical dashed line corresponds to $T_m = 3055 \pm11$~K, which is the value of melting temperature obtained in this work with the VNM technique, while the vertical dotted line corresponds to $T_m=2847\pm15$~K, which corresponds to the melting temperature obtained with the TPC method. Experimental and previously assessed values are shown for comparison, as well as results previously obtained with BMH potential.}
\label{fig:delta_h}
\end{figure}

The enthalpy of fusion ($\Delta H_{f}$) is defined as the heat required for a crystal to transition from a solid state to a liquid state at constant pressure. Due to the high temperatures required to run the experiments and the several technical issues related to the experimental setup, the enthalpy of fusion for CaO is usually difficult to obtain experimentally. This is the reason why the only available experimental data concerning this quantity are not direct measurements, but come from thermodynamic assessments\cite{Deffrennes2020} and the NIST-JANAF Tables value\cite{Chase1998}, which is estimated from MgO. Besides the uncertainty, this experimental assessments place the value of CaO enthalpy of fusion around 80~kJ/mol, which is the value we take as comparison as the best experimental estimate in this work.\\
To estimate $\Delta H_{f}$, we employed both ab-initio molecular dynamics (AIMD) and classical molecular dynamic (CMD) with the MLIP to calculate the caloric curve at ambient pressure for the crystal phase and the liquid phase. Starting with classical MD, we start from a 9$\times$9$\times$9 supercell of CaO, which correspond to 5832 atoms in total, at 300~K and then we gradually increase the temperature up until we observe homogeneous melting. Once we obtained the liquid phase, we gradually cool down the system, decreasing the temperature starting at 4000~K up until we observe an apparent recrystallization of the system. Both the heating and the cooling of the system are stepwise, with every point of the curve being the result of a NPT simulation 50,000 steps long, with a timestep of 0.5~fs and a temperature step of 100~K. The enthalpy for each temperature is calculated as the average enthalpy value during the last 40,000 steps of each simulation. \\
Regarding AIMD simulations, instead, the available computational power allowed us to run only a much smaller supercell of 3$\times$ 3$\times$ 3 (216 atoms in total) for 1,000 steps of 1~fs each. We perform all the simulations in the canonical ensemble (NVT) with a velocity-rescaling thermostat, at different temperatures to try to cover the entire temperature range of interest, especially around the expected $T_m$ of the system. To account for the solid branch of the enthalpy, we run AIMD simulations on the crystal phase at 300~K, 500~K, 1000~K, 1500~K, 2000~K, 3000~K, 3100~K, 3150~K and 3200~K; to account for the liquid branch, instead, we run AIMD simulations on the liquid phase at 3000~K, 3100~K, 3150~K, 3200~K, 3600~K and 4000~K. \\
The results are shown in Figure \ref{fig:delta_h}. 

Both curves are plotted with respect to a reference temperature of 300~K and are truncated one step before the change of phase occurs, that is at 3300~K for the crystal phase and at 2000~K for the liquid phase. The plot shows that the results obtained employing the MLIP are in excellent agreement with both the available experimental data and the AIMD calculations, following the same trend and the same slope. The same cannot be stated for the results previously obtained\cite{menescardi2024} with a BMH empirical potential, which, despite the difference between the solid and the liquid curves gives a similar result, tend to part ways from the experimental and the AIMD data as the temperature increases, resulting in a completely different slope. We can say that employing a MLIP significantly improved the accuracy of these calculations, bringing the MD simulations results to be almost superimposed to the enthalpy data tabulated by NIST-JANAF\cite{Chase1998} all the way up to 4000~K. AIMD calculations are in good agreement too, despite some fluctuations probably due to the small size of the supercell and the much shorter simulation time we were able to perform due to the limited computational power available.\\
The enthalpy of fusion ($\Delta H_f$) is calculated as the difference between the enthalpy of the liquid phase and the enthalpy of the solid phase at the melting temperature ($T_m$) of the system. In Fig. \ref{fig:delta_h} we show the results for the calculation of the enthalpy of fusion. The dashed vertical line is the reference melting temperature as calculated with the VNM technique ($T_m=3055\pm11$~K), while the dotted vertical line corresponds to the melting temperature obtained with TPC method ($T_m=2847\pm14$~K). We obtain $\Delta H_f$=73.46~kJ/mol at a melting temperature of 3055~K and $\Delta H_f$=71.59~kJ/mol at a melting temperature of 2847~K, which are in agreement with previously  obtained values, both from experimental assessments and simulations. The enthalpy of fusion, in fact, depends on the melting temperature it is referring to, which is very debated for this particular system, with the consequence that the values of $\Delta H_f$ reported in literature strongly depend on the $T_m$ taken as reference. For example, previous thermodynamic assessments by Defrennes et al.\cite{Deffrennes2020} obtained a $\Delta H_f=80\pm16$ kJ/mol, assuming the $T_m$ of CaO to be 3222~K, as experimentally reported by Manara et al.\cite{Manara2014}. Even though the enthalpy of fusion obtained in this work falls in the range of the value obtained by Defrennes et al., a better comparison would be between the values of the entropy of fusion ($\Delta S_f$) which is calculated as $\Delta S_f=\frac{\Delta H_f}{T_m}$ and can be directly confronted with the tabulated experimental values. In table \ref{tab:deltaH}, we show the resulting values of $\Delta S_f$.

\begin{table*}\begin{center}
\renewcommand{\arraystretch}{1.3}
\begin{tabular}{cccccc}
\hline\hline
\textbf{$\Delta S_f$ ($J\cdot mol^{-1}\cdot K^{-1}$)} & \textbf{Method} & \textbf{Reference} \\
\hline
25.14 &  CMD (MLIP) @ 2847 K & This work \\
24.05   & CMD (MLIP) @ 3055 K& This work \\
27.77   & AIMD @ 3100 K & This work \\
25.35   & AIMD @ 3150 K & This work \\
26.21   & CMD (BMH potential) & Menescardi et al.\cite{menescardi2024}\\
29.21   & therm. assessment @ 3222 K& Belmonte et al.\cite{Belmonte2017a}\\
25 $\pm$ 5   & estimated from BeO & Gurvich et al. \cite{gurvich1994}\\
25.06   & estimated from MgO & NIST-JANAF Tables\cite{Chase1998}\\
\hline\hline
\end{tabular}
\caption{Values of $\Delta S_f$ obtained in this work compared to previous CMD results and available experimental data.}\label{tab:deltaH}
\end{center}\end{table*}

\begin{figure}[h!]
\centering
    \includegraphics[width=\linewidth]{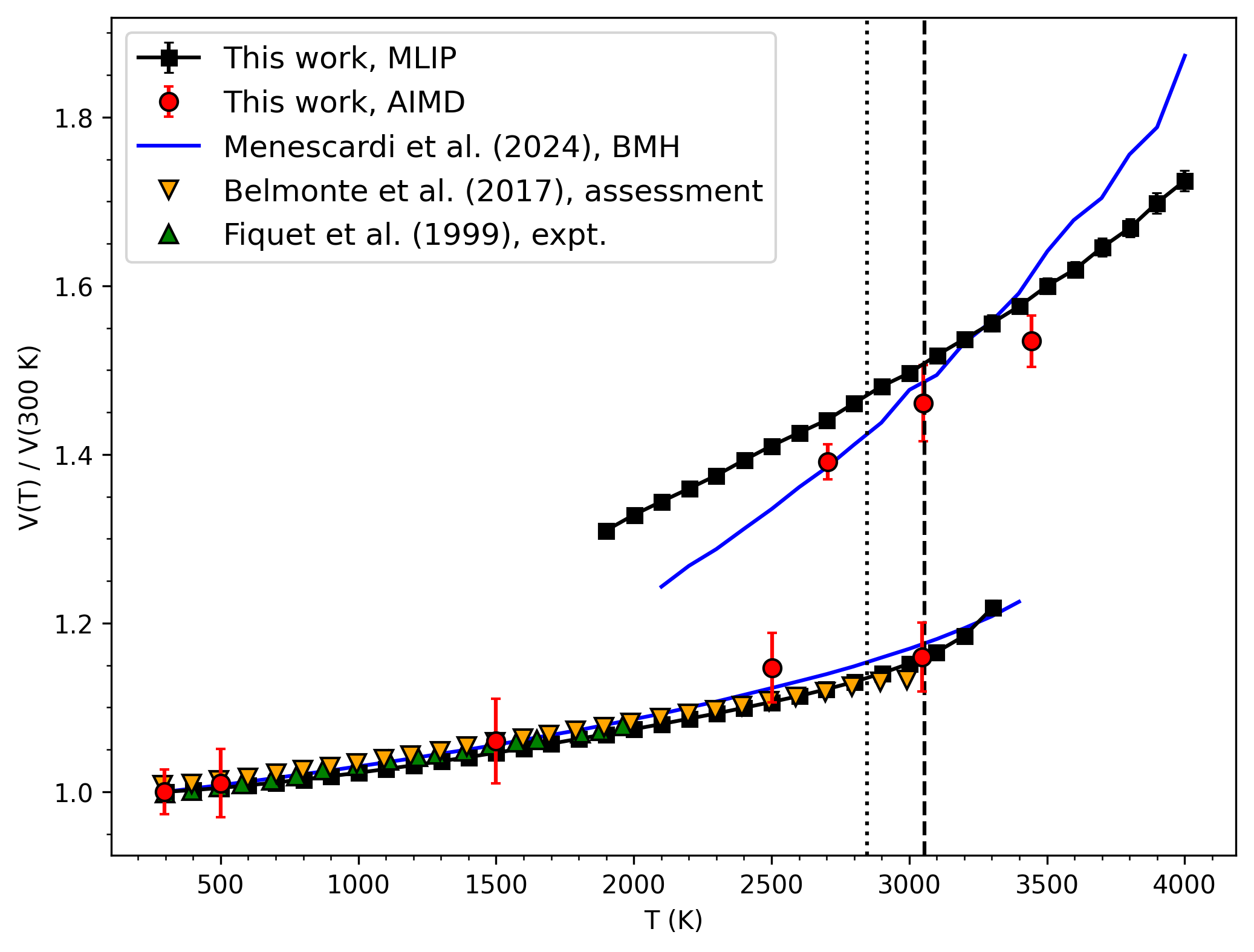}
    \caption{Relative volume thermal expansion as a function of temperature. \revised{Error bars represent the standard deviation of the volume during the production phase of each simulation (last 40,000 of 50,000 steps for MLIP-MD; last 1700 of 2000 steps for AIMD).} The lower branch is referred to the solid phase while the upper branch corresponds to the liquid phase. The vertical dashed line corresponds to $T_m = 3055 \pm11$~K, which is the value of melting temperature obtained in this work with the VNM technique, while the vertical dotted line corresponds to $T_m=2847\pm15$~K, which is the melting temperature obtained employing the TPC technique. Experimental data\cite{Fiquet1999, Belmonte2017a} and data previously obtained with BMH potential\cite{menescardi2024} are shown for comparison. }
\label{fig:VvsT}
\end{figure}

The obtained $\Delta S_f$ obtained in this work are very close to the values tabulated by NIST-JANAF\cite{Chase1998} and by Gurvich et al.\cite{gurvich1994}, which are estimated from MgO and BeO, respectively. The agreement is also strong with AIMD calculations and the previous calculations employing a BMH potential\cite{menescardi2024}, while it slightly underestimates the value from Belmonte et al.\cite{Belmonte2017a}, which was obtained from  an \textit{ab initio}-assisted assessment assuming a melting temperature of 3222~K.  We reported two values from AIMD calculation because, even if 3100~K is the closest calculation to the predicted $T_m=3055\pm11$, the calculations made at the very close temperature of 3150~K are closer to the experimental and theoretical curve, and thus more reliable, as shown in Figure \ref{fig:delta_h}. \\
In addition to the caloric curve, we report the relative volume thermal expansion as a function of temperature at ambient pressure, in Figure \ref{fig:VvsT}. 

Our results are in very good agreement with the reference data: the volume ratio $V(T)/V(300\,\mathrm{K})$ underlying Fig.~\ref{fig:VvsT} agrees to within 1\% with the experimental data.~\cite{Fiquet1999} over the 300-2000~K interval considered, and to within 0.5\% with the thermodynamic assessment~\cite{Belmonte2017a} up to $\sim$2200~K, diverging above $\sim$2400~K to reach $\sim$2.2\% by 3000~K as the melting temperature is approached. \revised{To further quantify this agreement, we computed the volumetric thermal expansion coefficient.}

\begin{equation}
    \alpha_V = \frac{1}{V}\left(\frac{\partial V}{\partial T}\right)_P .
\end{equation}

\revised{Over the 300-2000~K range shown in Fig. \ref{fig:VvsT}, we obtain a mean thermal expansion coefficient of $\bar{\alpha}_V \approx 4.3\times10^{-5}~\mathrm{K^{-1}}$, within 6\% of the value derived from the experimental data from Fiquet et al.~\cite{Fiquet1999} ($\bar{\alpha}_V \approx 4.6\times10^{-5}~\mathrm{K^{-1}}$) over the same interval. Comparison with the ab-initio-assisted thermodynamic assessment of Belmonte et al.~\cite{Belmonte2017a}, which extends to higher temperature, shows excellent agreement up to $\sim$2200~K ($\bar{\alpha}_V \approx 4.4\times10^{-5}~\mathrm{K^{-1}}$, average difference $1.4\%$), followed by a marked, monotonically increasing departure above $\sim$2400~K, consistent with the onset of anharmonic behavior due to premelting captured by our MD simulation but not resolvable by the low-order polynomial form of the thermodynamic assessment.} \\
We observe that the slope of the liquid branch of the thermal expansion differ significantly from the one previously obtained\cite{menescardi2024} with a BMH potential. To verify the validity of MLIP potential, we run AIMD simulation in the isobaric ensemble (NPT) at different temperatures to compare DFT data to the results already obtained. The AIMD simulation are performed on a 3$\times$3$\times$3 supercell for 2000 steps of 1 fs each. The values reported are the average cell volume over the last 1700 steps, after 300 steps equilibration, at different temperatures and ambient pressure.\\
While the AIMD values lie at slightly lower absolute volumes in the liquid branch - an effect likely attributable to the constraints of the small 216-atom supercell and limited simulation time - they replicate the flatter slope of the MLIP potential, as shown in figure \ref{fig:VvsT}. This observation confirms that DFT-calculated thermal expansion curve for the liquid phase ought to have a smaller slope than the corresponding one calculated with BMH potential. One reasoning for these very different slopes could be that the rigid repulsive wall of the BMH potential is responsible for the large thermal volumetric expansion of the liquid at high temperature, while AIMD and MLIP potential account partially for the oxygen polarization due to the surrounding cations, which gives rise to soft repulsive interaction at short range. This behavior for BMH potential is well known, as it was already reported to underestimate DFT-calculated volume in CaO\cite{Alvares2020} and to have a higher slope in the liquid branch of the curve in the case of MgO\cite{arkhipin2024}.\\
Besides the differences in slope, the estimation of the volume jump at the melting temperature is still similar, as we obtain a volume expansion of $\sim$29\% at both reference melting temperatures considered, which aligns with values obtained for other ionic systems\cite{Zykova2005a} and values previously assessed for CaO\cite{Sun2010}. The magnitude of the volume expansion at melting is also consistent with the density decrease obtained from independently computed number density profiles shown in Sec.\ref{sec:TPC}, providing an internal validation of the simulation framework.
\subsection{High-Pressure Melting Curve}
\begin{figure}[h]
\centering
    \includegraphics[width=\linewidth]{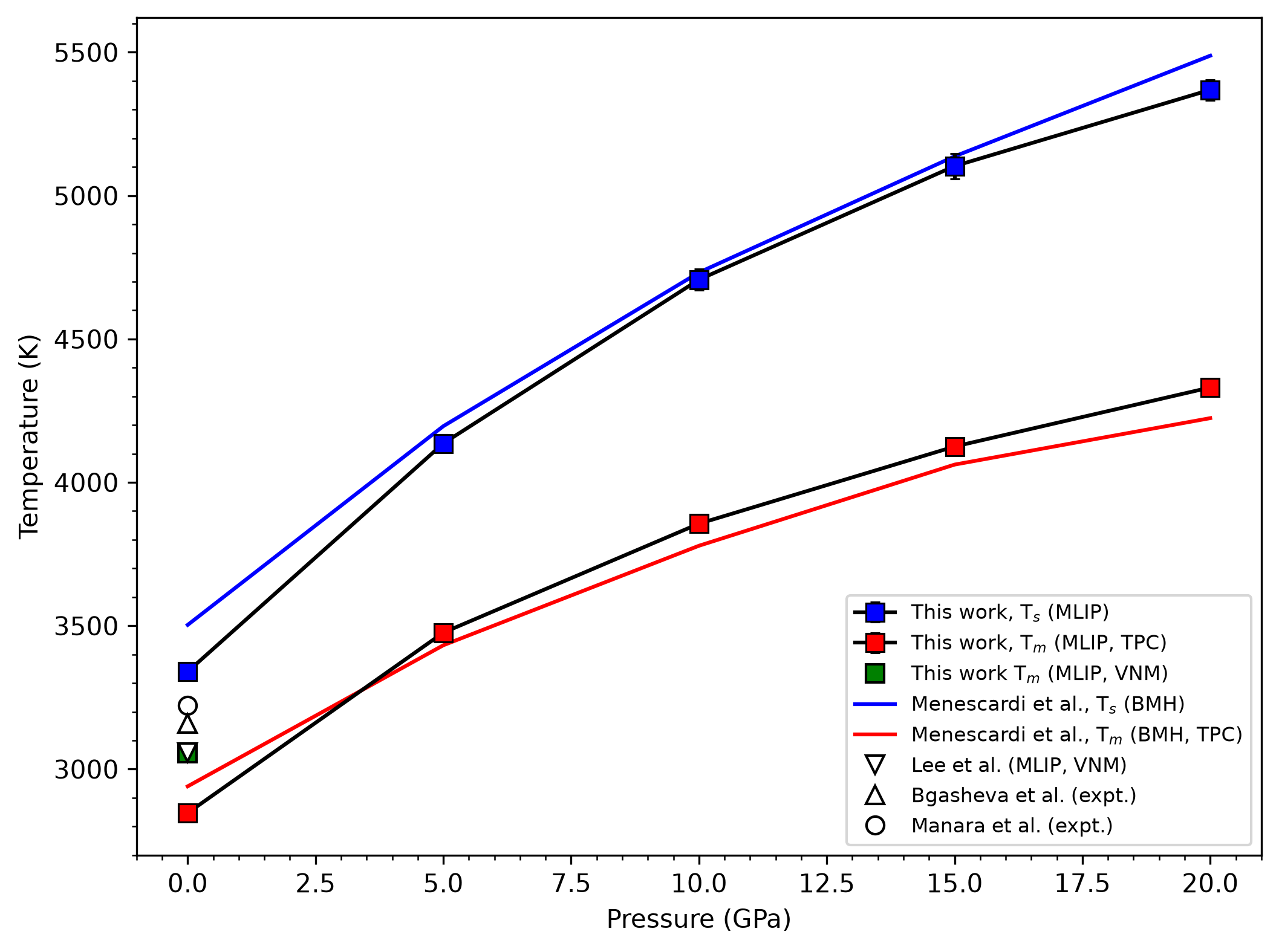}
    \caption{High pressure melting curve of CaO ($T_m$, red squares) and thermal instability temperature curve ($T_s$, blue squares) calculated with MLIP potential. The corresponding solid lines refer to the curves previously obtained with BMH potential\cite{menescardi2024}, while other symbols at ambient pressure refer to all the most recent values of $T_m$ reported in literature, both from experiments\cite{Bgasheva2021, Manara2014} and MD simulations\cite{Lee2022}. \revised{Error bars on $T_s$ represent the standard deviation of the temperature within the melting transition window. Error bars on $T_m$ represent the standard deviation of the instantaneous temperature over the NPH production run.}}
\label{fig:HPcurve}
\end{figure}

Here we present the result for the high pressure melting curve for CaO up to 20 GPa of pressure. \revised{While VNM yielded a value numerically closer to the latest experimental data at ambient pressure (see Section~\ref{sec:VNM}),} this technique relies on the presence of increasingly bigger voids, which are bound to undermine the stability and eventually cause the collapse of the crystal structure when external pressure is applied. For this reason, we employ the TPC method to calculate the high-pressure melting curve for the crystal. \\
Very few attempts were made in previous works to calculate the high pressure melting curve for CaO. The attempt made by Sun et al.\cite{Sun2010} in 2010, which relies on the assumption that the high pressure melting curve would maintain a 30\% overheat regardless of the pressure, was proven inaccurate by our previous calculations employing BMH potential\cite{menescardi2024}, in which we employed the TPC technique on much smaller supercells. \\
In this work, to obtain the high pressure $T_m$ curve, we employ the TPC technique  employing a 6$\times$6$\times$60 supercell (17,280 atoms), following the same procedure proposed for ambient conditions\cite{asadi2015} at different pressures, starting at 0 GPa, which corresponds to the results already shown in section \ref{sec:TPC}, and increasing with a step of 5 GPa up to 20 GPa, which is the highest pressure the MLIP can reliably handle, since it was trained with structures in that pressure range. In order to get the thermal instability temperature ($T_s$) curve, instead, we perform a $NPT$ simulation on a perfect  11$\times$11$\times$11 supercell  (10,648 atoms), increasing the temperature at a rate of $10^{12}$~K/s for 0.5 ns at different pressures, with the same pressure step as the TPC calculations performed to obtain the melting curve. Since the crystal, in this case, has no defects, it melts at the thermal instability limit, which is affected by a significant overheating, which for these systems is usually 20\%-25\%\cite{Jin2001}.\\
Figure \ref{fig:HPcurve} shows the results obtained in this work compared to the available experimental data at ambient pressure and the curves obtained with BMH potential in our previous work.

The plot shown the high-pressure melting curve rapidly flattening as the pressure increases, following the slope of the curve already obtained in 2024 using BMH potential, and thus confirming the result. The $T_m$ and the $T_s$ high-pressure curves, in fact, do not have the same slope, as the overheating ratio, defined as $\eta=\frac{T_s}{T_m}-1$ increases with increasing pressure, going form an overheat of about 17\% at ambient pressure to a maximum of 24\% at P=20 GPa. All the calculated $\eta$ values and the corresponding temperatures are shown in table \ref{tab:press}.

\begin{table*}\begin{center}
\renewcommand{\arraystretch}{1.3}
\begin{tabular}{cccc}
\hline\hline
\textbf{P (GPa)} & \textbf{$T_m$ (K)} & \textbf{$T_s$ (K)} & \textbf{$\eta$ (\%)} \\
\hline
0   & 2847 $\pm$ 15 & 3340 $\pm$ 21 & 17.3 \\
5   & 3476 $\pm$ 22 & 4136 $\pm$ 31 & 19.0 \\
10   & 3856 $\pm$ 22 & 4707 $\pm$ 37 & 22.1 \\
15   & 4125 $\pm$ 25 & 5103 $\pm$ 44 & 23.7 \\
20   & 4331 $\pm$ 30 & 5369 $\pm$ 36 & 24.0 \\
\hline\hline
\end{tabular}
\caption{Values of $T_m$ and $T_s$ obtained in this work, with the corresponding percentage overheating ratio ($\eta$)}\label{tab:press}
\end{center}\end{table*}

Following the rationale discussed by Gomez et al.\cite{Gomez2005}, who extensively studied the same phenomenon on Lennard-Jones crystals, we fit both curves into empirical equations of the form $T(P)=aP^{b}+c$ to obtain the following equations:
\begin{eqnarray}
    T_m(P)=284.4P^{0.601}+2844.1 \nonumber\\
    T_s(P)=331.6P^{0.589}+3337.2 \label{eq:fitting}
\end{eqnarray}

From these equations it can be observed that the exponent $b$ has very similar value for both curves, which indicates that $T_m$ and $T_s$ curves follow the same scaling laws. The pre-factor  $a$, instead, is significantly bigger for the $T_s$ curve, which reflects a bigger slope when compared to the $T_m$ curve, confirming a significantly different trend as pressure is applied. This behavior was explained\cite{Gomez2005} by observing that the lack of nucleation sites in the perfect crystal, from which the $T_s$ curve arises, requires a higher thermal activation barrier, which increases as the pressure increases. This dramatic increase in the activation energy does not arise in presence of a solid-liquid interface, since the external pressure applied acts mostly on the more compressible liquid phase. These results reflect those already obtained with BMH potential\cite{menescardi2024}, ultimately confirming the novel result then proposed.\\
\revised{As a further consistency check, we verified that the initial slope of the $T_m(P)$ curve from Eq.~\ref{eq:fitting} is compatible with the Clausius-Clapeyron relation evaluated at $T_m=2847$~K, $P=0$.}

\begin{equation}
    \left.\frac{dT_m}{dP}\right|_{P=0} = \frac{\Delta V}{\Delta S},
\end{equation}

\revised{Using the independently computed volume jump at melting ($\Delta V \approx 29\%$, i.e.\ $\Delta V \approx 5.72~\mathrm{cm^3/mol}$ in absolute molar terms) and the entropy of fusion already reported in Table~ \ref{tab:deltaH} ($\Delta S_f = 25.14~\mathrm{J\,mol^{-1}\,K^{-1}}$), the Clausius-Clapeyron relation predicts a slope of $\approx228~\mathrm{K/GPa}$ at ambient pressure. The analytical derivative of the empirical fit in Eq.~\ref{eq:fitting} diverges as $P\to0$ and cannot be evaluated exactly at zero pressure, but, at a small but finite pressure ($P \approx 0.5$~GPa) the derivative of Eq.~\ref{eq:fitting} reaches $\approx225~\mathrm{K/GPa}$, matching the Clausius--Clapeyron estimate to within $\sim$1\%. While a strict comparison at $P=0$ is precluded by the functional form adopted, this close agreement in the  immediate vicinity of ambient pressure supports the thermodynamic consistency of the fitted high-pressure melting curve with the independently computed volume and entropy jumps at $T_m$.}
\section{Conclusions}
This study successfully implemented a Machine Learning Interatomic Potential (MLIP) to investigate the melting temperature and the high-temperature and high-pressure phase stability of calcium oxide (CaO). Employing the PANNA 2.0 framework with the LATTE descriptor, we achieved \textit{ab-initio} accuracy with the computational efficiency required for large-scale molecular dynamics simulations, that constitutes a substantial improvement over the empirical BMH potential used in our previous work.\\
We compared the two-phase coexistence (TPC) method and the void-nucleated melting (VNM) technique to calculate the melting temperature at ambient pressure, obtaining $T_m=2847\pm15$~K and $T_m=3055\pm11$~K respectively. \revised{These results are consistent with results previously obtained with the same techniques; while VNM yields a value numerically closer to the latest experimental determinations of $T_m$, neither method can be regarded as intrinsically more reliable than the other.}\\
Next, we calculated the enthalpy difference ($\Delta H$) curve, which was found to be perfectly consistent with available experimental data and with AIMD calculations, improving the results previously obtained with BMH potential. The corresponding enthalpies of fusion ($\Delta S_{f}$) were found to be $\Delta S_{f}=24.05~J\cdot mol^{-1}\cdot K^{-1}$ at $T_m=3055\pm11$~K and $\Delta S_{f}=25.14~J\cdot mol^{-1}\cdot K^{-1}$ at $T_m=2847\pm15$~K, which are both consistent with tabulated experimental data and previously assessed results. \\
Finally, using the TPC method, we calculated the high-pressure melting ($T_m(P)$) curve for CaO up to 20 GPa and compared it to the thermal instability limit ($T_s(P)$) curve. The results show that $T_m$ increases less steeply that $T_s$ as pressure increases, confirming the results already obtained with BMH potential and ultimately confirming that assuming a constant $\frac{T_s}{T_m}$ ratio, as sometimes done in literature\cite{Sun2010}, is not an accurate strategy.
\section*{CRediT authorship contribution statement}
\textbf{Francesca Menescardi:} Conceptualization, Methodology, Software, Investigation, Data curation, Formal analysis, Visualization, Writing - original draft.\\
\textbf{Stefano de Gironcoli:} Writing - review and editing, Supervision, Funding acquisition.
\section*{Data availability}
The dataset generated during the current study is fully available in the Zenodo repository at the following DOI: \hyperlink{https://doi.org/10.5281/zenodo.19814455}{https://doi.org/10.5281/zenodo.19814455}. \\
Representative input files and starting configurations of the MD calculations performed during the current study are available in the GitHub repository at the following link:\\ \hyperlink{https://github.com/fmenescardi/CaO-from-Neural-Network-Potentials}{https://github.com/fmenescardi/CaO-from-Neural-Network-Potentials}.
\section*{Declaration of generative AI and AI-assisted technologies in the manuscript preparation process}
During the preparation of this work the authors used NotebookLM in order to improve the language and formatting of the manuscript. After using this tool, the authors reviewed and edited the content as needed and take full responsibility for the content of the published article.
\section*{Acknowledgements}
This study was funded by the European Union - NextGenerationEU, Mission 4, Component 2, Inv.1.4, in the framework of the PNRR Project National Centre for HPC, Big Data and Quantum Computing (CN HPC) CN00000013; CUP G93C22000600001. SdG acknowledges support also from the European Commission through the MAX
Centre of Excellence for supercomputing applications (grant
numbers 10109337 and 824143)\\
High performance calculations were carried out thanks to resources provided by CINECA HPC.\\
The authors thank Dr. Davide Ceresoli for helpful suggestions and the useful discussions. 
\section*{Competing interests}
All authors declare no financial or non-financial competing interests.
%

\bibliography{references}

@article{Agrawal2003,
  title        = {Molecular dynamics study of the melting of nitromethane},
  author       = {Agrawal, Paras M. and Rice, Betsy M. and Thompson, Donald L.},
  year         = 2003,
  month        = nov,
  journal      = {The Journal of Chemical Physics},
  publisher    = {AIP Publishing},
  volume       = 119,
  number       = 18,
  pages        = {9617--9627},



}

@article{Alvares2020,
  title        = {Thermodynamics and structural properties of {CaO}: A molecular dynamics simulation study},
  author       = {Alvares, Cecilia M. S. and Deffrennes, Guillaume and Pisch, Alexander and Jakse, No{\"e}l},
  year         = 2020,
  month        = feb,
  journal      = {The Journal of Chemical Physics},
  publisher    = {AIP Publishing},
  volume       = 152,
  number       = 8,
  pages        = 084503,



}

@article{Belmonte2017a,
  title        = {Ab initio-assisted assessment of the \ce{CaO}-\ce{SiO2} system under pressure},
  author       = {Belmonte, D. and Ottonello, G. and Zuccolini, M. Vetuschi},
  year         = 2017,
  month        = dec,
  journal      = {Calphad},
  publisher    = {Elsevier BV},
  volume       = 59,
  pages        = {12--30},



}

@article{Bgasheva2021,
  title        = {Laser-pulse melting of calcium oxide and some peculiarities of its high-temperature behavior},
  author       = {Bgasheva, Tatiana and Falyakhov, Timerkhan and Petukhov, Sergei and Sheindlin, Michael and Vasin, Andrey and Vervikishko, Pavel},
  year         = 2021,
  month        = mar,
  journal      = {Journal of the American Ceramic Society},
  publisher    = {Wiley},
  volume       = 104,
  number       = 7,
  pages        = {3461--3477},



}

@article{Csonka2009,
  title        = {Assessing the performance of recent density functionals for bulk solids},
  author       = {Csonka, G{\'a}bor I. and Perdew, John P. and Ruzsinszky, Adrienn and Philipsen, Pier H. T. and Leb{\`e}gue, S{\'e}bastien and Paier, Joachim and Vydrov, Oleg A. and {\'A}ngy{\'a}n, J{\'a}nos G.},
  year         = 2009,
  month        = apr,
  journal      = {Physical Review B},
  publisher    = {American Physical Society (APS)},
  volume       = 79,
  number       = 15,
  pages        = 155107,



}

@article{Giannozzi2009,
  title        = {{Quantum ESPRESSO}: a modular and open-source software project for quantum simulations of materials},
  author       = {Giannozzi, Paolo and Baroni, Stefano and Bonini, Nicola and Calandra, Matteo and Car, Roberto and Cavazzoni, Carlo and Ceresoli, Davide and Chiarotti, Guido L and Cococcioni, Matteo and Dabo, Ismaila and Dal Corso, Andrea and de Gironcoli, Stefano and Fabris, Stefano and Fratesi, Guido and Gebauer, Ralph and Gerstmann, Uwe and Gougoussis, Christos and Kokalj, Anton and Lazzeri, Michele and Martin-Samos, Layla and Marzari, Nicola and Mauri, Francesco and Mazzarello, Riccardo and Paolini, Stefano and Pasquarello, Alfredo and Paulatto, Lorenzo and Sbraccia, Carlo and Scandolo, Sandro and Sclauzero, Gabriele and Seitsonen, Ari P and Smogunov, Alexander and Umari, Paolo and Wentzcovitch, Renata M},
  year         = 2009,
  month        = sep,
  journal      = {Journal of Physics: Condensed Matter},
  publisher    = {IOP Publishing},
  volume       = 21,
  number       = 39,
  pages        = 395502,



}

@article{Giannozzi2017,
  title        = {Advanced capabilities for materials modelling with {Quantum ESPRESSO}},
  author       = {Giannozzi, P and Andreussi, O and Brumme, T and Bunau, O and Buongiorno Nardelli, M and Calandra, M and Car, R and Cavazzoni, C and Ceresoli, D and Cococcioni, M and Colonna, N and Carnimeo, I and Dal Corso, A and de Gironcoli, S and Delugas, P and DiStasio, R A and Ferretti, A and Floris, A and Fratesi, G and Fugallo, G and Gebauer, R and Gerstmann, U and Giustino, F and Gorni, T and Jia, J and Kawamura, M and Ko, H-Y and Kokalj, A and K{\"u}{\c{c}}{\"u}kbenli, E and Lazzeri, M and Marsili, M and Marzari, N and Mauri, F and Nguyen, N L and Nguyen, H-V and Otero-de-la-Roza, A and Paulatto, L and Ponc{\'e}, S and Rocca, D and Sabatini, R and Santra, B and Schlipf, M and Seitsonen, A P and Smogunov, A and Timrov, I and Thonhauser, T and Umari, P and Vast, N and Wu, X and Baroni, S},
  year         = 2017,
  month        = oct,
  journal      = {Journal of Physics: Condensed Matter},
  publisher    = {IOP Publishing},
  volume       = 29,
  number       = 46,
  pages        = 465901,



}

@article{Giannozzi2020,
  title        = {{Quantum ESPRESSO} toward the exascale},
  author       = {Giannozzi, Paolo and Baseggio, Oscar and Bonf{\`a}, Pietro and Brunato, Davide and Car, Roberto and Carnimeo, Ivan and Cavazzoni, Carlo and de Gironcoli, Stefano and Delugas, Pietro and Ferrari Ruffino, Fabrizio and Ferretti, Andrea and Marzari, Nicola and Timrov, Iurii and Urru, Andrea and Baroni, Stefano},
  year         = 2020,
  month        = apr,
  journal      = {The Journal of Chemical Physics},
  publisher    = {AIP Publishing},
  volume       = 152,
  number       = 15,
  pages        = {154105},



}

@article{Gomez2005,
  title        = {Pressure dependence of the melting mechanism at the limit of overheating in {Lennard-Jones} crystals},
  author       = {G{\'o}mez, L. and Gazza, C. and Dacharry, H. and Pe{\~n}aranda, L. and Dobry, A.},
  year         = 2005,
  month        = apr,
  journal      = {Physical Review B},
  publisher    = {American Physical Society (APS)},
  volume       = 71,
  number       = 13,
  pages        = {134106},



}

@article{Hong2013,
  title        = {Solid-liquid coexistence in small systems: A statistical method to calculate melting temperatures},
  author       = {Hong, Qi-Jun and van de Walle, Axel},
  year         = 2013,
  month        = sep,
  journal      = {The Journal of Chemical Physics},
  publisher    = {AIP Publishing},
  volume       = 139,
  number       = 9,



}

@article{Jin2001,
  title        = {Melting Mechanisms at the Limit of Superheating},
  author       = {Jin, Z. H. and Gumbsch, P. and Lu, K. and Ma, E.},
  year         = 2001,
  month        = jul,
  journal      = {Physical Review Letters},
  publisher    = {American Physical Society (APS)},
  volume       = 87,
  number       = 5,
  pages        = 055703,



}

@article{Lee2022,
  title        = {Ab initio construction of full phase diagram of \ce{MgO}-\ce{CaO} eutectic system using neural network interatomic potentials},
  author       = {Lee, Kyeongpung and Park, Yutack and Han, Seungwu},
  year         = 2022,
  month        = nov,
  journal      = {Physical Review Materials},
  publisher    = {American Physical Society (APS)},
  volume       = 6,
  number       = 11,
  pages        = {113802},



}

@article{Liang2018,
  title        = {Complete thermodynamic description of the \ce{Mg-Ca-O} phase diagram including the \ce{Ca-O}, \ce{Mg-O} and \ce{CaO-MgO} subsystems},
  author       = {Liang, Song-Mao and Schmid-Fetzer, Rainer},
  year         = 2018,
  month        = nov,
  journal      = {Journal of the European Ceramic Society},
  publisher    = {Elsevier BV},
  volume       = 38,
  number       = 14,
  pages        = {4768--4785},



}

@article{Lutsko1989,
  title        = {Molecular-dynamics study of lattice-defect-nucleated melting in metals using an embedded-atom-method potential},
  author       = {Lutsko, J. F. and Wolf, D. and Phillpot, S. R. and Yip, S.},
  year         = 1989,
  month        = aug,
  journal      = {Physical Review B},
  publisher    = {American Physical Society (APS)},
  volume       = 40,
  number       = 5,
  pages        = {2841--2855},



}

@article{Manara2014,
  title        = {On the melting behaviour of calcium monoxide under different atmospheres: A laser heating study},
  author       = {Manara, D. and B{\"o}hler, R. and Capriotti, L. and Quaini, A. and Bao, Z. and Boboridis, K. and Luzzi, L. and Janssen, A. and P{\"o}ml, P. and Eloirdi, R. and Konings, R.J.M.},
  year         = 2014,
  month        = jun,
  journal      = {Journal of the European Ceramic Society},
  publisher    = {Elsevier BV},
  volume       = 34,
  number       = 6,
  pages        = {1623--1636},



}

@article{Morris1994,
  title        = {Melting line of aluminum from simulations of coexisting phases},
  author       = {Morris, J. R. and Wang, C. Z. and Ho, K. M. and Chan, C. T.},
  year         = 1994,
  month        = feb,
  journal      = {Physical Review B},
  publisher    = {American Physical Society (APS)},
  volume       = 49,
  number       = 5,
  pages        = {3109--3115},



}

@article{Perdew2008,
  title        = {Restoring the Density-Gradient Expansion for Exchange in Solids and Surfaces},
  author       = {Perdew, John P. and Ruzsinszky, Adrienn and Csonka, G{\'a}bor I. and Vydrov, Oleg A. and Scuseria, Gustavo E. and Constantin, Lucian A. and Zhou, Xiaolan and Burke, Kieron},
  year         = 2008,
  month        = apr,
  journal      = {Physical Review Letters},
  publisher    = {American Physical Society (APS)},
  volume       = 100,
  number       = 13,
  pages        = {136406},



}

@article{Plimpton1995,
  title        = {Fast Parallel Algorithms for Short-Range Molecular Dynamics},
  author       = {Plimpton, Steve},
  year         = 1995,
  month        = mar,
  journal      = {Journal of Computational Physics},
  publisher    = {Elsevier BV},
  volume       = 117,
  number       = 1,
  pages        = {1--19},



}

@article{Sun2010,
  title        = {The high-pressure melting curve of \ce{CaO}},
  author       = {Sun, X.W. and Song, T. and Chu, Y.D. and Liu, Z.J. and Zhang, Z.R. and Chen, Q.F.},
  year         = 2010,
  month        = oct,
  journal      = {Solid State Communications},
  publisher    = {Elsevier BV},
  volume       = 150,
  number       = {37--38},
  pages        = {1785--1788},



}

@article{Thompson2022,
  title        = {{LAMMPS} -- a flexible simulation tool for particle-based materials modeling at the atomic, meso, and continuum scales},
  author       = {Thompson, Aidan P. and Aktulga, H. Metin and Berger, Richard and Bolintineanu, Dan S. and Brown, W. Michael and Crozier, Paul S. and in 't Veld, Pieter J. and Kohlmeyer, Axel and Moore, Stan G. and Nguyen, Trung Dac and Shan, Ray and Stevens, Mark J. and Tranchida, Julien and Trott, Christian and Plimpton, Steven J.},
  year         = 2022,
  month        = feb,
  journal      = {Computer Physics Communications},
  publisher    = {Elsevier BV},
  volume       = 271,
  pages        = 108171,



}

@article{Troullier1991,
  title        = {Efficient pseudopotentials for plane-wave calculations},
  author       = {Troullier, N. and Martins, Jos{\'e} Luriaas},
  year         = 1991,
  month        = jan,
  journal      = {Physical Review B},
  publisher    = {American Physical Society (APS)},
  volume       = 43,
  number       = 3,
  pages        = {1993--2006},



}

@article{Wang2023,
  title        = {Prediction of the melting curve and phase diagram for \ce{CaO} using newly developed interatomic potentials},
  author       = {Wang, Xin-Wei and Sun, Xiao-Wei and Song, Ting and Tian, Jun-Hong and Liu, Zi-Jiang},
  year         = 2023,
  month        = mar,
  journal      = {Vacuum},
  publisher    = {Elsevier BV},
  volume       = 209,
  pages        = 111717,



}

@article{Zykova2005a,
  title        = {Physics of solid and liquid alkali halide surfaces near the melting point},
  author       = {Zykova-Timan, T. and Ceresoli, D. and Tartaglino, U. and Tosatti, E.},
  year         = 2005,
  month        = oct,
  journal      = {The Journal of Chemical Physics},
  publisher    = {AIP Publishing},
  volume       = 123,
  number       = 16,
  pages        = {164701},



}

@article{Hoover1985,
    title={Canonical dynamics: Equilibrium phase-space distributions},
    volume={31},



    number={3},
    journal={Physical Review A},
    publisher={American Physical Society (APS)},
    author={Hoover, William G.},
    year={1985},
    month=mar,
    pages={1695--1697}
}

@book{Anderson1989,
    place={Boston, MA},
    title={Theory of the Earth},

    publisher={Blackwell Scientific Publications},
    author={Anderson, Don L.},
    year={1989},
    month={gen}
}

@article{Humphrey1996,
    title={{VMD}: Visual molecular dynamics},
    volume={14},



    number={1},
    journal={Journal of Molecular Graphics},
    publisher={Elsevier BV},
    author={Humphrey, William and Dalke, Andrew and Schulten, Klaus},
    year={1996},
    month=feb,
    pages={33--38}
}

@book{Chase1998,
  author = {Chase M},
  title = {NIST-JANAF Thermochemical Tables, 4th Edition},
  year = {1998},
  month = {1998-08-01},
  publisher = {American Institute of Physics},
  language = {en},
}

@book{Taylor1997,
  title={Cement Chemistry},
  author={Taylor, H.F.W.},

  lccn={98176503},

  year={1997},
  publisher={Emerald Publishing Limited}
}

@article{Belonoshko2006,
  title={Melting and critical superheating},
  volume={73},



  number={1},
  pages={012201},
  journal={Physical Review B},
  publisher={American Physical Society (APS)},
  author={Belonoshko, A. B. and Skorodumova, N. V. and Rosengren, A. and Johansson, B.},
  year={2006},
  month=jan
}

@article{Zhang2012, title={A comparison of methods for melting point calculation using molecular dynamics simulations}, volume={136}, number={14}, journal={The Journal of Chemical Physics}, publisher={AIP Publishing}, author={Zhang, Yong and Maginn, Edward J.}, year={2012}, month=apr, pages={144116} }

@article{Hong2022, title={Melting temperature prediction via first principles and deep learning}, volume={214}, journal={Computational Materials Science}, publisher={Elsevier BV}, author={Hong, Qi-Jun}, year={2022}, month=nov, pages={111684} }

@article{Fiquet1999, 
title={High-temperature thermal expansion of lime, periclase, corundum and spinel},
volume={27},



number={2},
journal={Physics and Chemistry of Minerals},
publisher={Springer Science and Business Media LLC},
author={Fiquet, G. and Richet, P. and Montagnac, G.},
year={1999}, month=dec,
pages={103--111}
}

@article{pellegrini2023panna2,
  author    = {Franco Pellegrini and Ruggero Lot and Yusuf Shaidu and Emine K{\"u}{\c{c}}{\"u}kbenli},
  title     = {{PANNA} 2.0: Efficient neural network interatomic potentials and new architectures},
  journal   = {The Journal of Chemical Physics},
  volume    = {159},
  number    = {8},
  year      = {2023},

}

@article{lot2020panna,
  author    = {Ruggero Lot and Franco Pellegrini and Yusuf Shaidu and Emine K{\"u}{\c{c}}{\"u}kbenli},
  title     = {Panna: Properties from artificial neural network architectures},
  journal   = {Computer Physics Communications},
  volume    = {256},
  pages     = {107402},
  year      = {2020},

}

@misc{pellegrini2024latte,
  author    = {Franco Pellegrini and Stefano de Gironcoli and Emine K{\"u}{\c{c}}{\"u}kbenli},
  title     = {{LATTE}: an atomic environment descriptor based on Cartesian tensor contractions},
  note      = {arXiv preprint arXiv:2405.08137},
  year      = {2024},

}

@article{behler2007,
  author    = {J{\"o}rg Behler and Michele Parrinello},
  title     = {Generalized Neural-Network Representation of High-Dimensional Potential-Energy Surfaces},
  journal   = {Physical Review Letters},
  volume    = {98},
  number    = {14},
  pages     = {146401},
  year      = {2007},
  month     = {April},

}

@article{menescardi2024,
  title={Melting behavior of \ce{CaO} at high temperature and pressure: a molecular dynamics study},
  author={Menescardi, Francesca and Ceresoli, Davide and Belmonte, Donato},
  journal={The Journal of Physical Chemistry C},
  volume={128},
  number={43},
  pages={18498--18508},
  year={2024},
  publisher={ACS Publications}
}

@article{zou2020,
  title={Investigation on the efficiency and accuracy of methods for calculating melting temperature by molecular dynamics simulation},
  author={Zou, YangChun and Xiang, ShiKai and Dai, ChengDa},
  journal={Computational Materials Science},
  volume={171},
  pages={109156},
  year={2020},
  publisher={Elsevier}
}

@article{wang2024,
  title={A comprehensive investigation on the accuracy and efficiency of methods for melting temperature calculation using molecular dynamics simulations},
  author={Wang, Xinwei and Yang, Mengxin and Gai, Xiaoqian and Sun, Yibo and Cao, Bohan and Chen, Jiajin and Liang, Min and Tian, Fubo and Li, Liang},
  journal={Journal of Molecular Liquids},
  volume={395},
  pages={123924},
  year={2024},
  publisher={Elsevier}
}

@article{asadi2015,
  title={Two-phase solid--liquid coexistence of Ni, Cu, and Al by molecular dynamics simulations using the modified embedded-atom method},
  author={Asadi, Ebrahim and Zaeem, Mohsen Asle and Nouranian, Sasan and Baskes, Michael I},
  journal={Acta Materialia},
  volume={86},
  pages={169--181},
  year={2015},
  publisher={Elsevier}
}

@article{giorgino2014,
  title={Computing 1-D atomic densities in macromolecular simulations: The density profile tool for VMD},
  author={Giorgino, Toni},
  journal={Computer Physics Communications},
  volume={185},
  number={1},
  pages={317--322},
  year={2014},
  publisher={Elsevier}
}

@article{wang2024construction,
  title={Construction and application of deep learning potential for CaO under high pressure},
  author={Wang, Xinwei and Liu, Zi-Jiang and Feng, Jin-Shan and Chen, Meng-Ru and Li, Liang and Sun, Xiao-Wei and Tian, Fubo},
  journal={Computational Materials Science},
  volume={244},
  pages={113154},
  year={2024},
  publisher={Elsevier}
}

@article{gurvich1994,
  title={Elements \ce{B}, \ce{Al}, \ce{Ga}, \ce{In}, \ce{Tl}, \ce{Be}, \ce{Mg}, \ce{Ca}, \ce{Sr}, \ce{Ba}, and Their Compounds. \ce{CRC Press} and \ce{Begell House}. Part One. Methods and Computation. VIII+ 707 pp. Part Two. Tables},
  author={Gurvich, LV and Veyts, IV and Tab Alcock, CB and Iorish, VS},
  journal={Thermodynamic Properties of Individual Substances},
  volume={3},
  pages={XV+--448},
  year={1994}
}

@article{arkhipin2024,
  title={A Molecular Dynamics Simulation Study of Crystalline and Liquid \ce{MgO}},
  author={Arkhipin, Anatoly S and Pisch, Alexander and Uspenskaya, Irina A and Jakse, No{\"e}l},
  journal={Ceramics},
  volume={7},
  number={3},
  pages={1187--1203},
  year={2024},
  publisher={MDPI}
}

@article{Kirkwood1935,
    author = {Kirkwood, John G.},
    title = {Statistical Mechanics of Fluid Mixtures},
    journal = {The Journal of Chemical Physics},
    volume = {3},
    number = {5},
    pages = {300-313},
    year = {1935},
    month = {05},
    issn = {0021-9606},
    doi = {10.1063/1.1749657},
    url = {https://doi.org/10.1063/1.1749657},
    eprint = {https://pubs.aip.org/aip/jcp/article-pdf/3/5/300/18788537/300_1_online.pdf},
}

@book{FrenkelSmit2023,
  author    = {Frenkel, Daan and Smit, Berend},
  title     = {Understanding Molecular Simulation: From Algorithms to Applications},
  edition   = {3rd},
  publisher = {Academic Press},
  year      = {2023},
  month     = jul,
  isbn      = {9780323902922},
  language  = {English}
}

@article{Leite2019,
  title={Nonequilibrium free-energy calculations of fluids using LAMMPS},
  author={Leite, Rodolfo Paula and de Koning, Maurice},
  journal={Computational Materials Science},
  volume={159},
  pages={316--326},
  year={2019},
  publisher={Elsevier}
}

@article{Freitas2016,
    title = {Nonequilibrium free-energy calculation of solids using LAMMPS},
    journal = {Comput. Mater. Sci.},
    volume = {112},
    pages = {333 - 341},
    year = {2016},
    author = {Freitas, Rodrigo and Asta, Mark and de Koning, Maurice}
}

@article{behler2016,
  title={Perspective: Machine learning potentials for atomistic simulations},
  author={Behler, J{\"o}rg},
  journal={The Journal of chemical physics},
  volume={145},
  number={17},
  year={2016},
  publisher={AIP Publishing}
}

@article{unke2021,
  title={Machine learning force fields},
  author={Unke, Oliver T and Chmiela, Stefan and Sauceda, Huziel E and Gastegger, Michael and Poltavsky, Igor and Schutt, Kristof T and Tkatchenko, Alexandre and Muller, Klaus-Robert},
  journal={Chemical reviews},
  volume={121},
  number={16},
  pages={10142--10186},
  year={2021},
  publisher={ACS Publications}
}

@article{butler2018,
  title={Machine learning for molecular and materials science},
  author={Butler, Keith T and Davies, Daniel W and Cartwright, Hugh and Isayev, Olexandr and Walsh, Aron},
  journal={Nature},
  volume={559},
  number={7715},
  pages={547--555},
  year={2018},
  publisher={Nature Publishing Group UK London}
}

@article{deringer2017,
  title={Machine learning based interatomic potential for amorphous carbon},
  author={Deringer, Volker L and Cs{\'a}nyi, G{\'a}bor},
  journal={Physical Review B},
  volume={95},
  number={9},
  pages={094203},
  year={2017},
  publisher={APS}
}

@article{artrith2012,
  title={High-dimensional neural network potentials for metal surfaces: A prototype study for copper},
  author={Artrith, Nongnuch and Behler, J{\"o}rg},
  journal={Physical Review B--Condensed Matter and Materials Physics},
  volume={85},
  number={4},
  pages={045439},
  year={2012},
  publisher={APS}
}

@article{jain2021,
  title={Machine learning for metallurgy III: A neural network potential for Al-Mg-Si},
  author={Jain, Abhinav CP and Marchand, Daniel and Glensk, Albert and Ceriotti, M and Curtin, WA},
  journal={Physical Review Materials},
  volume={5},
  number={5},
  pages={053805},
  year={2021},
  publisher={APS}
}
\end{document}